\documentclass[copyright,creativecommons]{eptcs}

\usepackage{multicol}
\usepackage{amsmath}
\usepackage{amssymb}

\usepackage{color}
\usepackage{xspace}
\usepackage{graphicx}
\usepackage{multirow}

\usepackage{amsthm}

\usepackage{tikz}
\usetikzlibrary{patterns,decorations.pathreplacing}
\usetikzlibrary{arrows}

\newcommand{\LeftEqNo}{\let\veqno\@@leqno}

\newcommand{\mso}{\mathit{MSO}}
\newcommand{\fo}{\mathit{FO}}

\newcommand{\ignore}[1]{}

\newtheorem{lemma}{Lemma}
\newtheorem{theorem}{Theorem}
\newtheorem{corollary}{Corollary}
\newtheorem{example}{Example}
\newtheorem{definition}{Definition}
\newtheorem{proposition}{Proposition}

\begin{document}

\thispagestyle{plain} 
\pagestyle{plain}     

%
\title{On the Path-Width of Integer Linear Programming}



\author{
Constantin Enea \qquad\qquad Peter Habermehl
\institute{LIAFA, University of Paris Diderot and CNRS\\75205 Paris 13\\France}
\email{\quad \{cenea, Peter.Habermehl\}@liafa.univ-paris-diderot.fr}
\and
Omar Inverso \qquad\qquad Gennaro Parlato
\institute{ School of  Electronics and Computer Science\\University of Southampton\\United Kingdom}
\email{\quad \{oi2c11,gennaro\}@ecs.soton.ac.uk}
}
\def\titlerunning{On the Path-Width of Integer Linear Programming}
\def\authorrunning{C. Enea, P. Habermehl, O. Inverso \& G. Parlato}

\maketitle

\makeatletter
\begin{abstract}
We consider the feasibility problem of \textit{integer linear programming} (ILP). We show that solutions of any ILP instance can be naturally represented by an FO-definable class of graphs. For each solution there may be many graphs representing it. However, one of these graphs is of path-width at most $2n$, where $n$ is the number of variables in the instance. Since FO is decidable on graphs of bounded path-width, we obtain an alternative decidability result for ILP. The technique we use underlines a common principle to prove decidability which has previously been employed for automata with auxiliary storage. We also show how this new result links to automata theory and program verification. 
\end{abstract}

\section{Introduction}
Alur and Madhusudan in \cite{Alur} have proposed nested words as a natural graph representation of runs of pushdown automata (PDA). A run is a sequence of moves which relate consecutive configurations of the PDA. A move is represented by a node, and nodes are linked through a linear order capturing the sequence of moves in the run. Further, nodes corresponding to matching push and pop moves are also linked together through (nested) matching edges. Thus, nested words  naturally reflect the semantics of PDA.

This concept of representing runs with graphs has been extended to other classes of automata with multiple stacks and queues. For example, runs of multi-stack PDA can be represented as multiply-nested words, i.e. nested words with a nested relation for each stack. Similarly, runs of distributed automata can be represented with graphs. A distributed automaton consists of a finite number of PDAs communicating through unbounded queues. A natural graph representation for a run is composed of a finite number of nested words, each representing an execution of a single PDA, with additional edges modelling queues: a node representing the action of sending a message is linked to the corresponding node representing the action of receiving that message. 

A surprising result by Madhusudan and Parlato shows that those graph representations straightforwardly lead to uniform decision procedures for several problems on these automata. In~\cite{tree-width-popl},  it is shown that the emptiness problem for PDAs as well as several restrictions of multi-stack PDAs and distributed automata is decidable, as the class of graphs representing the runs of these automata has bounded tree-width, and furthermore it is definable in monadic second-order logic (MSO). Thus, checking the existence of an accepting run of those automata is
equivalent to the satisfiability of the MSO formula charactering runs on the class of graphs of bounded tree-width.
 The tree-width of a graph is a parameter that tells how close to a tree a graph is \cite{Bodlaender}. The problem of MSO satisfiability on graphs is undecidable in general, but decidable on the class of bounded tree-width graphs~\cite{courcelle,Seese91}.

Although this is a mathematical reduction from the emptiness problem for automata to MSO satisfiability on graphs, the novelty here is not the reduction itself. In fact, since the problem is decidable, one could first solve it and then write an MSO formula that is satisfiable on graphs of tree-width 1 if and only if the problem admits a positive answer. In contrast, the {\em principle} outlined in~\cite{tree-width-popl} is that a natural graph representation that logically captures the semantics of these automata--not containing algorithmic insights--is sufficient for decidability.

Among the problems that have been shown decidable using this principle we have: (1) state reachability problem \cite{tree-width-popl}, model-checking of LTL \cite{ahmedLTL,scopedFSTTCS}, and {\em generalised} LTL~\cite{salvatoreLOGIC} for various restrictions of multi-stack PDA \cite{boundedCS,phase,faouziORDERED,salvatoreSCOPED}, and (2) the reachability problem~\cite{tree-width-popl,Heussner} for subclasses of distributed automata that communicate through unbounded queues~\cite{gennaroTACAS08,muschollQUEUE}.
The surprising aspect is that the new proofs are uniform and radically different from the ones previously proposed in the literature which are specifically crafted using different techniques on a case-by-case basis. This strengthens the intuition that a common principle governs the decidability of (those) problems. 
In general, the above principle could be lifted to decision problems. Although it may not be always applicable, it is interesting to establish its generality or limits by looking at other decidability results known in the literature.



In this paper, we consider the {\em feasibility problem} for {\em integer linear programming} (ILP, for short) that asks whether, given a finite set $I$ of linear constraints, there is an assignment of its variables such that all the constraints are satisfied\footnote{W.l.o.g., we suppose that the variables are interpreted as  positive integers and that $I$ contains only equalities.}. We show that the decidability principle based on bounded tree-width graph representations applies to the ILP feasibility problem in a stronger sense as described below.

As a {\bf first contribution} we give a natural graph representation for the solutions of an instance $I$ of ILP. The nodes of the graph represent a unary encoding of the solution, i.e. each node is labelled with exactly one variable of $I$, and the number of nodes with the same label is the value of the corresponding variable in the solution. The edges are used to enforce the constraints of the system. For simplicity, consider a system with only one constraint, where each variable is associated with one coefficient. Depending on the sign of this coefficient each variable can contribute to the overall value of the constraint by either increasing or decreasing it. Each node will have a number of edges equal to the absolute value of the corresponding coefficient. We use edges to pair nodes whose corresponding coefficients have different signs. Thus, a graph with well-matched nodes is a solution. In case of multiple constraints, we reiterate the above mechanism for each constraint individually, labelling the edges with the constraint represented. Since multiple ``matchings'' are possible for the same solution, a solution may have several of those graphs representing it. We prove that the class of graphs representing the solutions of an instance $I$ can be defined in first-order logic. See Figure~\ref{fig:ex_graph} for an example of a solution for a two-constraints system.

In general, the class of graphs representing all solutions may have unbounded path-width. We show that, for any solution, there always exists a graph representing it of {\em path-width} at most $2n$, where $n$ is the number of variables of $I$, and this constitutes the {\bf second contribution} of the paper. 
The path-width of a graph measures its closeness to a path (rather than a tree, as for tree-width).  
This provides us with a restriction of the decidability principle outlined above for the case of ILP, where bounded path-width is already sufficient as opposed to the general case where the tree-width needs to be bounded. 


As a {\bf last contribution} we define, for each ILP instance $I$, a finite state automaton $A_I$ over the alphabet of $I$'s variables, such that the Parikh image \cite{parikh} of $A_I$ is exactly the set of all solutions of $I$. This construction relies on the proof of bounded path-width we provide. Furthermore, this automaton can also be seen as a Boolean program $P_I$ of size linear in the size of $I$ as opposed to the exponential size of $A_I$, such that $I$ is feasible iff a given location in $P_I$ is reachable. This gives a {\em symbolic} alternative to solve ILP  using program verification tools. 

\medskip
\noindent
\emph{\bf Organization of the paper.} In Sec.~\ref{preliminaries}, we give basic definitions on graphs, tree-width, MSO on graphs, and the feasibility problem of ILP. In Sec.~\ref{graphrepresentation}, we present the graph representation for ILP solutions, and give its FO characterisation. In Sec.~\ref{boundedpathwidth}, we give the bounded path-width theorem, and in Sec.~\ref{automata} we describe the automata for ILP. We conclude with some remarks and future work in Sec.~\ref{conclusions}.

\medskip
\noindent
\emph{\bf Related Work.} 
Many approaches are known for solving the ILP feasibility problem, based on, e.g., branch-and-bound~\cite{Land60}, the cutting-plane method~\cite{Gomory60}, the LLL algorithm~\cite{LLL82}, the Omega test~\cite{Pugh92},
finite-automata theory~\cite{Buchi60,Ganesh02,Wolper95}. The latter defines finite-automata representations for the set of solutions of an ILP instance but, differently from our approach, they are based on representing the binary encodings of the integers involved in the solutions. 
The exponential bound on the minimal solutions of an ILP instance~\cite{Papadimitriou81} implies that, for any feasible instance $I$, there is an exponential bound $B$, such that some (but not all) solutions have a graph representation of path-width bounded by $B$. 
We prove that there exists a bounded path-width graph representation for \emph{each} solution of an  instance $I$ and the bound depends only on the number of variables of $I$. 

\section{Preliminaries}\label{preliminaries}

\noindent
Given two integers $i$ and $j$ with $i\leq j$, we denote with $[i,j]$ the set of all integers $k$ such that  $i\leq k\leq j$. 
\bigskip

\noindent{\bf Monadic second-order logic on graphs:}
Fix two disjoint finite alphabets $\Sigma_V$ and $\Sigma_E$. A  \emph{$(\Sigma_V,\Sigma_E)$-labelled graph} is a structure $G=(V, E, \{V_a\}_{a\in\Sigma_V},\{E_b\}_{b \in \Sigma_E})$, where $V$ is a finite set of vertices, $E$ is a finite multi-set of (undirected) edges represented by unordered pairs of elements of $V$, for each $a\in\Sigma_V$, $V_a\subseteq V$ is a set of $a$-labelled vertices, and, for each $b\in\Sigma_E$, $E_b \subseteq E$ is a multi-set of $b$-labelled edges. 
When $\Sigma_V=\Sigma_E=\emptyset$, $G$ is called simply a graph.  Let $v,v'\in V$, and $\pi=v_0,v_1,\ldots, v_t$ be any  sequence of distinct  vertices of $G$ with $v=v_0$ and $v'=v_t$. A {\em path} in $G$ from $v$ to $v'$ is any sequence $\pi$ such that $\{v_{i-1},v_{i}\}\in E$, for every $i\in[1,t]$. In the rest of the paper, we denote any edge of the form $\{u,v\}$ simply with a pair $(u,v)$ with the meaning that it is an unordered pair.


We view graphs as logical structures,
where $V$ is the universe.
Each set of vertices $V_a$ is a unary relation on vertices and each multi-set of edges $E_b$ is a binary relation on vertices. {\em Monadic second-order logic} ($\mso$ for short) is nowadays the standard logic to express properties on these structures. 
We fix a countable set of first-order variables  
(denoted by lower-case symbols, e.g., $x,y$) and a countable set of second-order variables  (denoted by upper-case symbols, e.g. $X, Y$). 
The first-order, resp., second-order, variables are interpreted as vertices, resp., sets of vertices, in the graph.
An $\mso$ formula $\varphi$ is defined by the following grammar:

$$ 
\begin{array}{lcl}
\varphi &\triangleq &x\!=\!y \,\,\,\,\mid\,\,\,\, 
V_a(x)\,\,\,\,\mid\,\,\,\,  E_b(x,y)  \,\,\,\,\mid\,\,\,\, x\in X \,\,\,\,\mid\,\,\,\, 
\varphi \vee \varphi \,\,\,\,\mid\,\,\,\, \neg \varphi \,\,\,\,\mid\,\,\,\,
  \exists x. \varphi \,\,\,\,\mid\,\,\,\, 
  \exists X. \varphi 
\end{array}
  $$
where $a \in \Sigma_V$, $b\in \Sigma_E$, $x, y$ are first-order variables, and 
$X$ is a second-order variable.
The semantics of $\mso$ is defined as usual. First-order logic ($\fo$, for short) is the restriction of $\mso$ to formulas over first-order variables.

A class of $(\Sigma_V,\Sigma_E)$-labelled graphs $\mathcal{C}$ is \emph{$\mso$-definable}, resp., \emph{$\fo$-definable}, if there is an
$\mso$, resp., $\fo$, formula $\varphi$ such that $\mathcal{C}$ is exactly the class of $(\Sigma_V,\Sigma_E)$-labelled graphs that satisfy $\varphi$.

\bigskip
\noindent
{\bf Tree/path-width of graphs:}
A \emph{tree-decomposition} of a graph $G=(V,E)$ is a pair $(T, {\mathit bag})$, where $T = (N, \rightarrow)$ is a tree\footnote{ A tree $T$ is a graph having a special vertex called the {\em root} such that for every vertex $v$ of $T$ there is exactly one path from the root to $v$.}
and $bag: N \rightarrow 2^{V}$ is a function, that satisfies the following:

\begin{itemize}
  \item For every $v \in V$, there is a vertex $n \in N$ such that $v \in {\mathit bag}(n)$.
  \item For every edge $(u,v) \in E$, there is a vertex $n \in N$ such that $u,v \in {\mathit bag}(n)$.
  \item If $u \in ({\mathit bag(n)}\cap {\mathit bag}(n'))$, for vertices $n,n' \in N$, then for every $n''$ that lies on the unique undirected path from $n$ and $n'$ in $T$, $u \in {\mathit bag}(n'')$.
\end{itemize}

A \emph{path-decomposition} of a graph $G=(V,E)$ is a tree-decomposition $(T, {\mathit bag})$ such that $T$ is a linear graph (i.e., a tree with exactly two leaves). 

The \emph{width} of a tree/path-decomposition of $G$ is the size of the largest bag in it, minus one; i.e. $max_{n \in N}\{ |{\mathit bag}(n)| \} - 1$. The \emph{tree-width}, resp., \emph{path-width}, of a graph is the \emph{smallest} of the widths of any of its tree-decompositions, resp., path-decompositions. The notions of tree/path-decomposition and tree/path-width are extended to $(\Sigma_V,\Sigma_E)$-labelled graphs by ignoring vertex and edge labels.

\bigskip
\noindent
{\bf Satisfiability of $\mso$:}
The satisfiability problem for $\mso$ is undecidable in general 
but it is decidable when restricting the class of models to 
graphs of bounded tree/path-width. 

\medskip
\begin{theorem}[Seese~\cite{Seese91}]
The problem of checking, given $k \in \mathbb{N}$ and $\varphi \in \mso$ over $(\Sigma_V,\Sigma_E)$-labelled graphs, whether there is a $(\Sigma_V,\Sigma_E)$-labelled graph $G$ of tree-width at most $k$ that satisfies $\varphi$, is decidable.
\end{theorem}

\ignore{
Note that the above certainly does not imply that satisfiability of $\mso$ is decidable on \emph{any} class of graphs of bounded tree-width (take a non-recursive class of linear-orders/words for a counter-example). However, an immediate corollary is that satisfiability of $\mso$ is also decidable on any $\mso$-definable class of graphs $\mathcal{C}$ of bounded tree-width (if $\varphi_C$ defines the class of graphs, and $\varphi$ is the $\mso$ formula, we can instantiate the above theorem for $\varphi_C \wedge \varphi$).
}
\medskip
\begin{corollary}
Let $\mathcal{C}$ be an $\mso$ definable class of $(\Sigma_V,\Sigma_E)$-labelled graphs. The problem of checking, given $k \in \mathbb{N}$ and an $\mso$-formula $\varphi$, whether there is a graph $G \in \mathcal{C}$ of tree-width at most $k$ that satisfies $\varphi$, is decidable.
\end{corollary}

\medskip
\noindent
{\bf Integer Linear Programming (ILP):} An \emph{ILP instance} is constituted by a set of equations of the form $A\vec{x}=\vec{b}$, where $A=(a_{j,i})_{j\in[1,m],i\in[1,n]}$ is a $m\times n$ matrix, $\vec{x}=(x_i)_{i\in[1,n]}$ is a vector of size $n$, $\vec{b}=(b_j)_{j\in[1,m]}$ is a vector of size $m$, and all elements of $A$ and $\vec{b}$ are integers~\footnote{We consider ILP instances in \emph{standard form}. ILP instances expressed as inequalities, i.e., $A\vec{x}\leq \vec{b}$, can be converted to standard form by introducing slack variables.}. The \textit{ILP feasibility problem}, asks to check whether there exists an integer vector $\vec{s}$ of size $n$ such that  $A\vec{s}=\vec{b}$ ($\vec{s}$ is called a \emph{solution} of $A\vec{x}=\vec{b}$). For the sake of simplicity, in this paper we only consider solutions composed of non-negative integers.

\section{Graph representation for ILP solutions} \label{graphrepresentation}

Given an ILP instance $A\vec{x}=\vec{b}$, we define the set of graphs ${\mathcal G}[A\vec{x}=\vec{b}]$ having the property that each graph in ${\mathcal G}[A\vec{x}=\vec{b}]$  represents a solution of $A\vec{x}=\vec{b}$. On the other hand, for every solution of $A\vec{x}=\vec{b}$ there is at least one graph in ${\mathcal G}[A\vec{x}=\vec{b}]$ representing it (but possibly more than one). Furthermore, we show that ${\mathcal G}[A\vec{x}=\vec{b}]$ is FO definable, which gives a polynomial time reduction from the  ILP feasibility problem to the satisfiability problem of FO.

We first give the intuition behind the graph representation of a solution, before we formalize and prove the results outlined above. Consider an ILP instance $A\vec{x}=\vec{b}$ with $\vec{x}=(x_1,x_2,\ldots,x_n)$. A graph $G$ in  ${\mathcal G}[A\vec{x}=\vec{b}]$, if any, has the following features. Each vertex of $G$ is labelled with an index from the set $[0,n]$, and the tuple $\vec{s}=(s_1,s_2,\ldots,s_n)$ is a solution of $A\vec{x}=\vec{b}$, where $s_i$ is the number of $G$ vertices labelled with variable index $i$. Intuitively, all vertices of $G$ labelled with $i$ give a unary representation of $s_i$. Furthermore, $G$ has a unique vertex labelled with $0$, which represents the vector $\vec{b}$. To impose that $\vec{s}$ is a solution of $A\vec{x}=\vec{b}$, $G$ is equipped with edges labelled with indices of constraints (each edge is labelled with a unique index). In order to satisfy the $j$-th constraint we impose that every vertex labelled with a variable index $i\in[1,n]$ is the end-point of $|a_{j,i}|$ edges labelled with $j$. Similarly, the unique vertex representing $\vec{b}$ is the end-point of $|b_j|$ edges labelled with $j$.
A vertex also comes with a sign for each constraint: for an $i$-labelled vertex $v$ and the $j$-th constraint (1) if $i\in[1,n]$ (it is a variable index) then $v$ has the same sign as $a_{j,i}$, otherwise (2)  $v$ is the unique vertex labelled with $0$, and has the opposite sign of $b_{j}$. All edges labelled with $j$ concern the $j$-th constraint. Thus,
we further impose that an edge labelled with $j$ is always incident to vertices with opposite signs. Intuitively, since the end-points of vertices represent the constants of the matrix $A$ in unary,  we can do the arithmetic related to each constraint by just matching these end-points (through edges). In fact, for a constraint $j$ each node labelled with $i\in[1,n]$ will contribute with   $|a_{j,i}|$ edges with the same sign of $a_{j,i}$. A similar argument holds for the node labelled with $0$. Therefore, imposing the matchings described above we make sure that $a_{j,1}\cdot x_1+\ldots+a_{j,n}\cdot x_n=b_j$ holds. Since the matchings are imposed for all constraints we have that $G$ faithfully represents a solution for all the linear constraints. It is worth noting that, we do not deliberately impose how  matchings are accomplished. Thus, the same solution $\vec{s}$ may have several graphs in  ${\mathcal G}[A\vec{x}=\vec{b}]$ representing it. We now provide an example to illustrate this intuition.



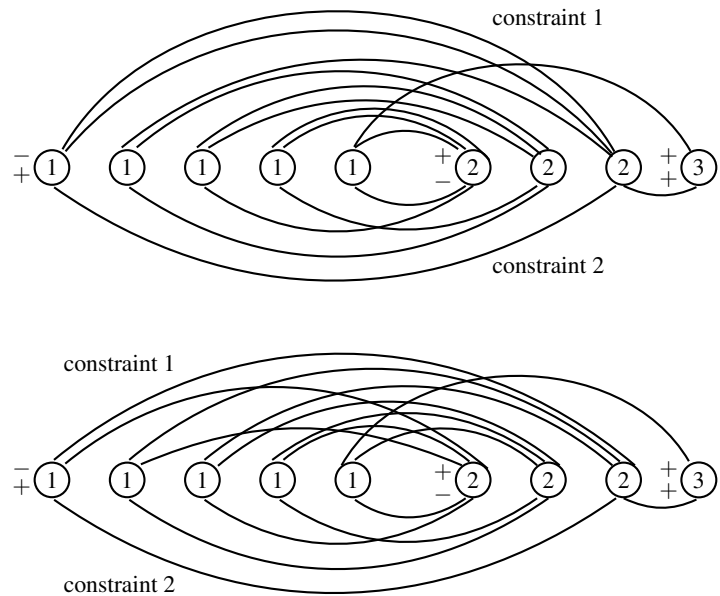
\begin{figure}[h]

\centering
%
{\footnotesize
\begin{tikzpicture}[-,>=stealth',shorten >=1pt,auto,node distance=1cm,
                            thick,main node/.style={circle,draw,inner sep=2},box/.style={circle,draw,dotted}]
	\node[main node] (D) at (0,0) {$1$};
	\node[above of=D,yshift=-.85cm,xshift=-4mm](p1){$-$};
	\node[above of=D,yshift=-1.1cm,xshift=-4mm](p1){$+$};
	\node[main node,right of=D] (E) {$1$};
	\node[main node,right of=E] (F) {$1$};
	\node[main node,right of=F] (G) {$1$};
	\node[main node,right of=G] (H) {$1$};
	\node[main node,right of=H,node distance=1.6cm] (A)  {$2$};
	\node[above of=A,yshift=-.85cm,xshift=-4mm](p1){$+$};
	\node[above of=A,yshift=-1.2cm,xshift=-4mm](p1){$-$};
	\node[main node,right of=A] (B) {$2$};
	\node[main node,right of=B] (C) {$2$};
	\node[main node,right of=C] (I) {$3$};
	\node[above of=I,yshift=-.85cm,xshift=-4mm](p1){$+$};
	\node[above of=I,yshift=-1.15cm,xshift=-4mm](p1){$+$};

	\node[above of=A,yshift=-2.3cm,xshift=10mm](p1){constraint 2};
	\node[above of=A,yshift=1.0cm,xshift=10mm](p1){constraint 1};

	\path
		 (C) edge  [bend right=50] node[above] {} (D) 
		 (C) edge  [bend right=60] node[above] {} (D) 
		 ([xshift=-1.5mm,yshift=-1mm]C.north) edge  [bend right=40] node[above] {} ([xshift=-1mm,yshift=0mm]E.north) 
		 
		 ([yshift=0mm,xshift=0mm]B.north) edge  [bend right=40] node[above] {} ([xshift=1mm]E.north) 
		 ([xshift=-1mm,yshift=-.7mm]B.north) edge  [bend right=40] node[above] {} ([xshift=-1mm]F.north) 
		 ([xshift=-2mm,yshift=-1mm]B.north) edge  [bend right=33] node[above] {} ([xshift=.5mm]F.north) 
		 ([yshift=1mm,xshift=-.3mm]A.east) edge  [bend right=45] node[above] {} ([xshift=-1mm]G.north) 
		 ([xshift=-1mm]A.north) edge  [bend right=38] node[above] {} ([xshift=.5mm]G.north) 
		 ([xshift=-2mm]A.north) edge  [bend right=36] node[above] {} (H.north) 
		 (I) edge  [bend right=60] node[above] {} (H.north) 
%
		 ([xshift=-1mm]C.south) edge  [bend left=35] node[above] {} (D.south) 
		 (C.south) edge  [bend right=25] node[above] {} (I.south) 
		 (B.south) edge  [bend left=35] node[above] {} (E.south) 
		 ([xshift=-1.5mm,yshift=0mm]B.south) edge  [bend left=35] node[above] {} (G.south) 
		 ([xshift=-1mm]A.south) edge  [bend left=35] node[above] {} (H.south) 
		 (A.south) edge  [bend left=35] node[above] {} (F.south) 

	;
\end{tikzpicture}
}
%
%

{\footnotesize
\begin{tikzpicture}[-,>=stealth',shorten >=1pt,auto,node distance=1cm,
                            thick,main node/.style={circle,draw,inner sep=2},box/.style={circle,draw,dotted}]

	\node[main node] (D) at (0,0) {$1$};
	\node[above of=D,yshift=-.85cm,xshift=-4mm](p1){$-$};
	\node[above of=D,yshift=-1.1cm,xshift=-4mm](p1){$+$};
	\node[main node,right of=D] (E) {$1$};
	\node[main node,right of=E] (F) {$1$};
	\node[main node,right of=F] (G) {$1$};
	\node[main node,right of=G] (H) {$1$};
                            
	\node[main node,right of=H,node distance=1.6cm] (A)  {$2$};
	\node[above of=A,yshift=-.9cm,xshift=-4mm](p1){$+$};
	\node[above of=A,yshift=-1.2cm,xshift=-4mm](p1){$-$};
	\node[main node,right of=A] (B) {$2$};
	\node[main node,right of=B] (C) {$2$};
	\node[main node,right of=C] (I) {$3$};
	\node[above of=I,yshift=-.85cm,xshift=-4mm](p1){$+$};
	\node[above of=I,yshift=-1.15cm,xshift=-4mm](p1){$+$};

	\node[above of=A,yshift=-2.38cm,xshift=-47mm](p1){constraint 2};
	\node[above of=A,yshift=0.55cm,xshift=-47mm](p1){constraint 1};

	\path
		 ([xshift=1.5mm,yshift=-.5mm]C.north) edge  [bend right=40] node[above] {} (D.north) 
		 ([xshift=-.5mm]C.north) edge  [bend right=40] node[above] {} (E.north) 
		 ([xshift=-1.5mm,yshift=-.5mm]C.north) edge  [bend right=40] node[above] {} (F.north) 
		 
		 ([xshift=-.5mm]B.north) edge  [bend right=37] node[above] {} ([xshift=-1mm]G.north) 
		 ([xshift=1.5mm,yshift=-.5mm]B.north) edge  [bend right=40] node[above] {} (F) 
		 ([xshift=-1.5mm,yshift=-.5mm]B.north) edge  [bend right=40] node[above] {} (H.north) 
		 ([xshift=-.5mm]A.north) edge  [bend right=40] node[above] {} ([xshift=.5mm]G.north) 
		 ([yshift=1.5mm,xshift=-.5mm]A.east) edge  [bend right=40] node[above] {} ([xshift=1.5mm,yshift=-.5mm]D.north) 
		 ([xshift=-1.5mm,yshift=-.5mm]A.north) edge  [bend right=23] node[above] {} ([xshift=1.5mm,yshift=-.5mm]E.north) 
		 ([xshift=-1.5mm,yshift=-.5mm]H.north) edge  [bend left=60] node[above] {} (I) 

		 ([xshift=-1mm]C.south) edge  [bend left=35] node[above] {} (D.south) 
		 (C.south) edge  [bend right=25] node[above] {} (I.south) 
		 (B.south) edge  [bend left=35] node[above] {} (E.south) 
		 ([xshift=-1.5mm,yshift=0mm]B.south) edge  [bend left=35] node[above] {} (G.south) 
		 ([xshift=-1mm]A.south) edge  [bend left=35] node[above] {} (H.south) 
		 (A.south) edge  [bend left=35] node[above] {} (F.south) 

%
%
	;
\end{tikzpicture}
}
%
%
	\caption{Two graph representations for the solution $x_1=5$, $x_2=3$, $x_3=1$ of $-2\, x_1 + 3\, x_2 + x_3 = 0$ and $x_1 -2\, x_2 + x_3 = 0$. The edges above, resp., below, the vertices  correspond to the first, resp, the second, equation. The signs attached to the vertices are the signs of the corresponding coefficients in the two constraints. 
The vertex labelled by $0$ is omitted because it has no incident edges. 
	}
	\label{fig:ex_graph}
\end{figure}

\begin{example}\label{ex:graphs}
The two graphs in Figure~\ref{fig:ex_graph} represent the solution $x_1=5$, $x_2=3$, $x_3=1$ of the ILP instance $-2\, x_1 + 3\, x_2 + x_3 = 0$ and $x_1 -2\, x_2 + x_3 = 0$.  
\end{example}

\begin{definition}\label{ILPgraphs}
Let $A\vec{x}=\vec{b}$ be an ILP instance. ${\mathcal G}[A\vec{x}=\vec{b}]$ is the set of all graphs 
$G=(V,E,\{V_i\}_{i\in[0,n]},$ $\{E_j\}_{j\in[1,m]})$, where: 
\begin{enumerate}
\setlength\itemsep{1mm}
	\item $V$ is a finite set of vertices and $\{V_i\}_{i\in[0,n]}$ defines a partition of $V$, i.e., for any $i\neq i'\in [0,n]$, $V_i\cap V_{i'}=\emptyset$ and $\bigcup_{i\in [0,n]} V_i=V$, and $|V_0|=1$;
	\item $E$ is a finite multi-set of edges and $\{E_j\}_{j\in[1,m]}$ defines a partition of $E$, i.e., for any $j\neq j'\in [1,m]$, $E_j\cap E_{j'}=\emptyset$ and $\bigcup_{j\in [1,m]} E_j=E$;
	\item if $(v,v')\in E_j$ with $v\in V_{i}$ and $v'\in V_{i'}$, then the signs of $a_{j,i}$ ($-b_j$ if $i=0$) and $a_{j,i'}$ ($-b_j$ if $i=0$) are different;
	\item $|\{(v,v')\in E_j\mid v\in V_0\}|=|b_j|$ and for any $i\in [1,n]$ and $v\in V_{i}$, $|\{(v,v')\in E_j\}|=|a_{j,i}|$.
\end{enumerate}
\medskip

%
\end{definition}

Next, we show that every graph in ${\mathcal G}[A\vec{x}=\vec{b}]$ defines a solution of $A\vec{x}=\vec{b}$ and vice-versa. Let $sol:{\mathcal G}[A\vec{x}=\vec{b}]\rightarrow \mathbb{N}^n$ be a function that associates to every graph $G$ in ${\mathcal G}[A\vec{x}=\vec{b}]$ a vector of natural numbers representing the number of vertices labelled with $i$, for each $i\in [1,n]$, i.e., $sol(G)=(|V_1|,\ldots,|V_n|)$.

\begin{proposition}\label{prop:graphs}
The image of the function $sol:{\mathcal G}[A\vec{x}=\vec{b}]\rightarrow \mathbb{N}^n$ is exactly the set of all solutions of $A\vec{x}=\vec{b}$.
\end{proposition}
\begin{proof}
Let $A\vec{x}=\vec{b}$ be an ILP instance and $G=(V,E,\{V_i\}_{i\in[0,n]},\{E_j\}_{j\in[1,m]})$ be a graph in ${\mathcal G}[A\vec{x}=\vec{b}]$. We show that for every $j\in [1,m]$, $sol(G)=(|V_1|,\ldots,|V_n|)$ is a solution of the equation $a_{j,1}\cdot x_1+\ldots +a_{j,n}\cdot x_n = b_j$. Let $a_{j,0}=-b_j$. The set of indices $[0,n]$ can be partitioned in two sets $\{p_1,\ldots,p_s\}$ and  $\{n_1,\ldots,n_t\}$ s.t. 
for every $k\in[1,s]$, $a_{j,p_k}$ is positive and for every $k\in[1,t]$, $a_{j,n_k}$ is negative. By definition, all the edges of $G$ labelled by $j$ are between a vertex in $V_{p_1}\cup\ldots\cup V_{p_s}$ and a vertex in $V_{n_1}\cup\ldots\cup V_{n_t}$. Also, for every $i\in [0,n]$, the degree of every vertex in $V_i$ equals $|a_{j,i}|$ and thus the number of edges labelled by $j$ can be written as both
\[
|V_{p_1}|\cdot a_{j,p_1} +\ldots + |V_{p_s}|\cdot a_{j,p_s}\mbox{ and }|V_{n_1}|\cdot |a_{j,n_1}| +\ldots + |V_{n_t}|\cdot |a_{j,n_t}|,
\]
which proves that $sol(G)$ is a solution of $a_{j,1}\cdot x_1+\ldots +a_{j,n}\cdot x_n = b_j$.

For the reverse, we show that for every solution $\vec{s}=(s_i)_{i\in[1,n]}$ of $A\vec{x}=\vec{b}$, there exists a graph $G=(V,E,\{V_i\}_{i\in[0,n]},\{E_j\}_{j\in[1,m]})$ in ${\mathcal G}[A\vec{x}=\vec{b}]$ s.t. $sol(G)=\vec{s}$. Therefore, for every $i\in [1,n]$, the set $V_i$ consists of $s_i$ vertices. Then, for every equation $a_{j,1}\cdot x_1+\ldots a_{j,n}\cdot x_n = b_j$ we consider the partition of $[0,n]$ into $\{p_1,\ldots,p_s\}$ and  $\{n_1,\ldots,n_t\}$ exactly as above. We also consider that $a_{j,0}=-b_j$ and $s_0=1$. The fact that $\vec{s}$ is a solution implies that 
\[
s_{p_1} \cdot a_{j,p_1} +\ldots + s_{p_s}\cdot a_{j,p_s}=s_{n_1}\cdot |a_{j,n_1}| +\ldots + s_{n_t}\cdot |a_{j,n_t}|,
\]
which shows that it is possible to define a multi-set of edges $E_j$ satisfying the constraints in Definition~\ref{ILPgraphs}.
\end{proof}

Proposition~\ref{prop:graphs} implies that the feasibility of an ILP instance is reducible to the problem of checking the existence of a graph satisfying the properties in Definition~\ref{ILPgraphs}.
\medskip
\begin{proposition} \label{lemmaIII}
\medskip

\noindent An ILP instance $A\vec{x}=\vec{b}$ is feasible iff ${\mathcal G}[A\vec{x}=\vec{b}]$ is non-empty.
\end{proposition}

\medskip

\noindent The following result shows that the class of graphs ${\mathcal G}[A\vec{x}=\vec{b}]$ from Definition~\ref{ILPgraphs} is definable in first-order logic. 

\medskip

\begin{proposition}\label{prop:fo}
For any ILP instance $A\vec{x}=\vec{b}$, there exists a first-order logic formula $\Phi[A\vec{x}=\vec{b}]$ such that for any graph $G$, $G \in {\mathcal G}[A\vec{x}=\vec{b}]$ iff $G \models \Phi[A\vec{x}=\vec{b}]$.
\end{proposition}
\begin{proof}

%
%
The formula $\Phi[A\vec{x}=\vec{b}]$ is defined as the conjunction of the formulae $\mathit{VertexLabels}$, $\mathit{Opposite}$, and $\mathit{Degree}$, which express condition (1), (3), and (4) in Def.~\ref{ILPgraphs}, respectively.

The condition on the vertex labels is given by the following formula:
\[
\begin{array}{lcl}
\mathit{VertexLabels}&\triangleq& \forall u. V_0(u)\oplus V_1(u)\oplus\ldots\oplus V_n(u)\\[.5mm]
&&\land\,\exists v. V_0(v) \wedge \forall w,w'. \big( (V_0(w) \wedge V_0(w')) \rightarrow w = w' \big),
\end{array}
\]
where $\oplus$ is the exclusive disjunction.


The formula $\mathit{Opposite}$ is defined by:
\[
\mathit{Opposite}\triangleq\forall u,v. \bigwedge\limits_{j \in [1,m]} \Big( E_j(u,v) \rightarrow \mathit{opposite}_j(u,v) \Big)
\]
where $\mathit{opposite}_j(u,v)$ says that the coefficients of the variables $x_i$ and $x_{i'}$ that label $u$ and resp., $v$, in the $j$th constraint, have opposite signs. Formally, for any $j\in [1,m]$, let $pos_j$ be the set of $i$ such $a_{j,i}\geq 0$ together with $0$, if $b_j\geq 0$. Analogously, let $neg_j$ be the set of $i$ such $a_{j,i}< 0$ together with $0$, if $b_j< 0$. Then,
{\small
\[ \mathit{opposite_j}(u,v) \triangleq 
				\left(\bigvee\limits_{i \in \mathit{pos}_j} V_i(u) \wedge
                                  \bigvee\limits_{i \in \mathit{neg}_j} V_i(v) \right) 
                                  \vee
                      \left(\bigvee\limits_{i \in \mathit{neg}_j} V_i(u) \wedge
                            \bigvee\limits_{i \in \mathit{pos}_j} V_i(v) \right).
                            \]
}



To express the constraint on the number of incident edges in a vertex of the graph, we introduce predicates of the form $E_j^k(u,v)$ with $k\in\mathbb{N}^*$, which holds iff there are exactly $k$ edges labelled by $j$ between $u$ and $v$. Let $max$ be the maximum value in $A$ or $\vec{b}$, in absolute value.
Then, 
{\small
\[
\mathit{Degree}\triangleq \underbrace{\forall u,v. \bigwedge_{j\in[1,m]} E_j(u,v) \rightarrow \big(E_j^1(u,v)\oplus\ldots\oplus E_j^{max}(u,v)\big)}_{\psi_1}
\land
\underbrace{\forall u. \bigwedge\limits_{\substack{i\in[0,n]\\ j\in[1,m]}} degree_{i,j}}_{\psi_2}
\]}

\noindent
where the sub-formula $\psi_1$ expresses the fact that, for any $u$ and $v$, there exists exactly one predicate $E_j^k(u,v)$ which holds and $ degree_{i,j}$ in $\psi_2$ is defined by:
{\small
\[
degree_{i,j} \triangleq V_{i}(u) \rightarrow \bigvee\limits_{\substack{z \leq |a_{i,j}|\\t_1,\dots,t_z > 0\\ t_1+\dots+t_z=|a_{i,j}|}}
\left( \begin{array}{c}\exists u_1,\dots,u_z . distinct(u_1,\dots,u_z)\\ \land E_i^{t_1}(u,u_1) \\\land E_i^{t_2}(u,u_2)\\ \dots \\\land E_i^{t_z}(u,u_z)  \end{array} \right)
\]
}
Above, $\mathit{distinct}(u_1, \dots,u_z)$ is the conjunction of all $u_i\neq u_{i'}$ with $1\leq i\neq i'\leq z$.  
%
\end{proof}

\section{Bounded Path-width}\label{boundedpathwidth}

In this section we show that for each solution $\vec{s}$ of $A\vec{x}=\vec{b}$ 
there is always a path-like graph representation in ${\mathcal G}[A\vec{x}=\vec{b}]$.
More precisely, we show that for any solution of an ILP instance with $n$ variables, there is a graph representation of this solution of path-width $2n$. Thus, the decidability of the ILP feasibility problem  can be directly derived by the decidability of FO on the class of bounded path-width graphs.

\medskip
\begin{lemma}\label{lem:pw}
For each solution $\vec{s}$ of $A\vec{x}=\vec{b}$ with $n$ variables, 
there is a graph $G \in {\mathcal G}[A\vec{x}=\vec{b}]$ whose path-width 
is upper-bounded by $2n$.
\end{lemma}
\begin{proof}
For each solution $\vec{s}$, 
${\mathcal G}[A\vec{x}=\vec{b}]$
may contain several graphs $G$ with $\vec{s}=sol(G)$. 
All of those have the same number of vertices with the same label as well 
as the same number of edges with the same label. Here, we show that among all graphs in 
${\mathcal G}[A\vec{x}=\vec{b}]$ representing $\vec{s}$, 
there exists one whose path-width is bounded by $2n$.
Without loss of generality we assume that $\vec{s}$ is a reduced solution,
i.e. it is not a multiple of another solution and that $\vec{s}$ does not
contain $0$.

Let $\vec{s} = (s_1, \dots, s_n)$.
The proof is given by fixing $\vec{b}=\vec{0}$. 
In this case the path-width can be bounded by $2n-1$.
At the end we show how to generalize the proof to any $\vec{b}$.
Any graph $G\in {\mathcal G}[A\vec{x}=\vec{b}]$ with $sol(G)=\vec{s}$ 
has $s_i$ vertices labelled by $i$. 
We say that $G \in {\mathcal G}[A\vec{x}=\vec{b}]$, with $G = (V,E,\{V_i\}_{i\in[0,n]},\{E_j\}_{j\in[1,m]})$,
is in {\em special form} if it satisfies the following two conditions:
(1) there is a partition $\{V^k\}_{k \in [1,t]}$ of $V$
such that no two vertices of $V^k$ are labelled with the same
variable index (i.e. $|V^k \cap V_i| \leq 1$ for all $i,k$), and (2) there is a partition
$\{E^k\}_{k \in [1,t]}$ of $E$ such that for all $k$,
all edges in $E^1\cup\ldots\cup E^k$ only relate vertices in
$V^1\cup\ldots\cup V^k$.

We now define a path decomposition for $G$ in special form.
We say that a vertex $v$ in $V$ in a subgraph of $G$ is fully matched if 
the subgraph contains all edges of $G$ involving $v$.
The bags of the path decomposition of $G$ are given 
by the sequence $B_p^0,B_p^1, B_p^2, \dots, B_p^{t}$. 
Initially, $B_p^0 = \emptyset$. At each step $k$, 
$B_p^{k+1} = B_p^k \setminus \{ v \in B_p^{k}  \mid v \textrm{ is fully matched in }
(V^1\cup\cdots\cup V^k,E^1\cup\cdots \cup E^k)\} \cup V^{k+1}$.

It is clear that the linear graph whose vertices are the bags
$B_p^1, \dots,  B_p^k$ is a path decomposition for $G$.
The size of the bags depends on the particular partition of vertices and
edges.

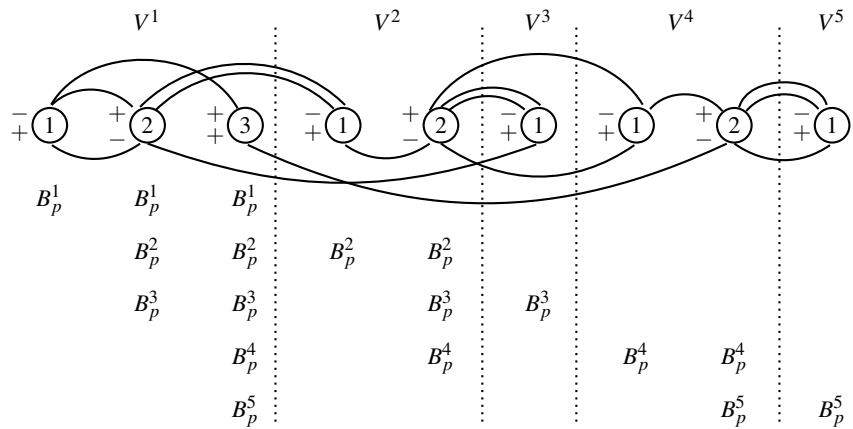
\begin{figure}[thb]
\centering
%
{\footnotesize
\begin{tikzpicture}[-,>=stealth',shorten >=1pt,auto,node distance=1.3cm,
                            thick,main node/.style={circle,draw,inner sep=2},box/.style={circle,draw,dotted}]
	\node[main node] (H) at (0,0) {$1$};
	\node[above of=H,node distance=1cm,yshift=-.85cm,xshift=-4mm](p1){$-$};
	\node[above of=H,node distance=1cm,yshift=-1.1cm,xshift=-4mm](p1){$+$};
	\node[below of=H,node distance=1cm](b11){$B_p^1$};

	\node[main node,right of=H] (A)  {$2$}; 
	\node[above of=A,node distance=1cm,yshift=-.85cm,xshift=-4mm](p1){$+$};
	\node[above of=A,node distance=1cm,yshift=-1.2cm,xshift=-4mm](p1){$-$};
	\node[below of=A,node distance=1cm](b12){$B_p^1$};
	\node[below of=b12,node distance=.7cm](b21){$B_p^2$};
	\node[below of=b21,node distance=.7cm](b31){$B_p^3$};
	\node[above of=A,node distance=1.4cm](p2){$V^1$};
	
	\node[main node,right of=A] (I) {$3$};
	\node[above of=I,node distance=1cm,yshift=-.85cm,xshift=-4mm](p1){$+$};
	\node[above of=I,node distance=1cm,yshift=-1.15cm,xshift=-4mm](p1){$+$};
	\node[below of=I,node distance=1cm](b13){$B_p^1$};
	\node[below of=b13,node distance=.7cm](b22){$B_p^2$};
	\node[below of=b22,node distance=.7cm](b32){$B_p^3$};
	\node[below of=b32,node distance=.7cm](b41){$B_p^4$};
	\node[below of=b41,node distance=.7cm](b53){$B_p^5$};
	\node[main node,right of=I] (G) {$1$};
	\node[above of=G,node distance=1cm,yshift=-.85cm,xshift=-4mm](p1){$-$};
	\node[above of=G,node distance=1cm,yshift=-1.1cm,xshift=-4mm](p1){$+$};
	\node[below of=G,node distance=1.7cm](b23){$B_p^2$};
	\node[above of=G,node distance=1.4cm,xshift=.6cm](p2){$V^2$};

	\node[main node,right of=G] (B) {$2$};
	\node[above of=B,node distance=1cm,yshift=-.85cm,xshift=-4mm](p1){$+$};
	\node[above of=B,node distance=1cm,yshift=-1.2cm,xshift=-4mm](p1){$-$};
	\node[below of=B,node distance=1.7cm](b24){$B_p^2$};
	\node[below of=b24,node distance=0.7cm](b33){$B_p^3$};
	\node[below of=b33,node distance=0.7cm](b41){$B_p^4$};

	\node[main node,right of=B] (F) {$1$};
	\node[above of=F,node distance=1cm,yshift=-.85cm,xshift=-4mm](p1){$-$};
	\node[above of=F,node distance=1cm,yshift=-1.1cm,xshift=-4mm](p1){$+$};
	\node[below of=F,node distance=2.4cm](b34){$B_p^3$};
	\node[above of=F,node distance=1.4cm](p2){$V^3$};

	\node[main node,right of=F] (E) {$1$};	
	\node[above of=E,node distance=1cm,yshift=-.85cm,xshift=-4mm](p1){$-$};
	\node[above of=E,node distance=1cm,yshift=-1.1cm,xshift=-4mm](p1){$+$};
	\node[below of=E,node distance=3.1cm](b42){$B_p^4$};
	\node[above of=E,node distance=1.4cm,xshift=.6cm](p2){$V^4$};

	\node[main node,right of=E] (C) {$2$};
	\node[above of=C,node distance=1cm,yshift=-.85cm,xshift=-4mm](p1){$+$};
	\node[above of=C,node distance=1cm,yshift=-1.2cm,xshift=-4mm](p1){$-$};
	\node[below of=C,node distance=3.1cm](b41){$B_p^4$};
	\node[below of=b41,node distance=0.7cm](b51){$B_p^5$};

	\node[main node,right of=C] (D) {$1$};
	\node[above of=D,node distance=1cm,yshift=-.85cm,xshift=-4mm](p1){$-$};
	\node[above of=D,node distance=1cm,yshift=-1.1cm,xshift=-4mm](p1){$+$};
	\node[below of=D,node distance=3.8cm](b52){$B_p^5$};
	\node[above of=D,node distance=1.4cm](p2){$V^5$};

	\path
		 (C) edge  [bend left=50] node[above] {} (D) 
		 (C) edge  [bend left=75] node[above] {} (D) 
		 
		 (C) edge  [bend right=50] node[above] {} (E) 
		 
		 ([xshift=-1.5mm,yshift=-.5mm]B.north) edge  [bend left=60] node[above] {} ([xshift=1mm]E.north) 
		 ([xshift=1mm,yshift=-.5mm]B.north) edge  [bend left=40] node[above] {} (F) 
		 (B.north) edge  [bend left=40] node[above] {} ([xshift=.5mm]F.north) 
		 ([xshift=1mm,yshift=-.5mm]A.north) edge  [bend left=45] node[above] {} ([xshift=-1mm,yshift=-.5mm]G.north) 
		 ([xshift=-1mm]A.north) edge  [bend left=45] node[above] {} ([xshift=.5mm]G.north) 
		 ([xshift=-2mm]A.north) edge  [bend right=45] node[above] {} (H.north) 
		 (I) edge  [bend right=60] node[above] {} (H.north) 
		 (C.south) edge  [bend right=35] node[above] {} (D.south) 
		 ([xshift=-1mm]C.south) edge  [bend left=25] node[above] {} (I.south) 
		 (B.south) edge  [bend right=35] node[above] {} (E.south) 
		 ([xshift=-1.5mm,yshift=0mm]B.south) edge  [bend left=35] node[above] {} (G.south) 
		 ([xshift=-1mm]A.south) edge  [bend left=35] node[above] {} (H.south) 
		 (A.south) edge  [bend right=20] node[above] {} (F.south) 
	;
	\draw[dotted] (3,1.3) -- (3,-4);
	\draw[dotted] (5.75,1.3) -- (5.75,-4);
	\draw[dotted] (7,1.3) -- (7,-4);
	\draw[dotted] (9.7,1.3) -- (9.7,-4);
\end{tikzpicture}
}
	\caption{A path decomposition of the first graph in Figure \ref{fig:ex_graph}
	}
	\label{fig:ex_graph2}
\end{figure}

In Figure \ref{fig:ex_graph2} we give the graph and its path decomposition computed by our algorithm explained below 
for the ILP instance in Example \ref{ex:graphs}.
The sets $V^1,\ldots,V^5$ are indicated by the dotted lines.
For each vertex, the bags $B_p^1,\ldots,B_p^5$ to which it belongs to are
indicated below it. Notice that the graph in the figure is isomorphic to the
first graph in Figure \ref{fig:ex_graph}.

In the following, we will define 
$\{V^k\}_{k \in [1,t]}$ and $\{E^k\}_{k \in [1,t]}$ 
such that the sizes of the bags are bounded by $2n$.
The idea is to pick the $V^k$ in a particular order such that it is always
possible to add edges making sufficiently many vertices fully matched 
which allows to drop them from the corresponding bag.
We will show that it is always
possible to have at most $2$ vertices labelled by the same variable in each
bag.

We will give now an auxiliary algorithm allowing us to define the partition
$\{V^k\}_{k \in [1,t]}$.
Consider two sets of counters $c_1, \dots,c_n$ and $r_1,\dots,r_m$ associated with the matrix $A$ in a way that we explain below. Let $s_{l} = max(s_1,\dots,s_n)$. Initially $\forall i. c_i = 0$. We define two possible actions on $c_i$: 

\begin{description}
\item[{\sc increase}$(i)$:] performs $c_i = c_i + s_l$;
\item[{\sc reduce}$()$:] performs $c_i = c_i - s_i$, $\forall i \in [1,n]$. 
\end{description}

When $c_i$ changes, all the counters $r_j$ are updated to $ \sum_{i}( (\textrm{number of } \textsc{increase}(i)) \cdot a_{j,i})$. It is clear that, if we perform exactly $s_i$ times $\textsc{increase}(i)$ for each $i \in [1,n]$ and $s_l$ times $\textsc{reduce}()$, all the counters reach zero. 
The meaning of the counters $c_i$ is 
purely functional to the algorithm we show below. 
The purpose of the counters $r_j$ is to tell how far (in the $j$-th constraint)  the solution is when the current assignment of the variable $x_i$ is set to the number of \textsc{increase($i$)}. When $r_j = 0$, the $j$-th constraint is satisfied.

Given the above mechanism, the counters $c_i$ and $r_j$ will range within a bounded interval if we use the following algorithm to determine the exact sequence of steps to perform:

\begin{enumerate}
\item \textsc{increase($i$)} while there is some $i$ such that $c_{i} < s_i$
\item \textsc{reduce()} and stop if $\forall i. c_{i} = 0$
\item  goto (1.)
\end{enumerate}

It is easy to see that 
for all counters $c_i$ we have $0 \leq c_i < 2 \cdot s_l$
and after reduce steps $0 \leq c_i \leq s_l$.
For the solution of the ILP instance of Example \ref{ex:graphs}
the sequence of counter values $(c_1,c_2,c_3)$ computed before and
after each of the five \textsc{reduce()} steps is $(0,0,0) \rightarrow \cdots 
(5,5,5)
\rightarrow_{\textsc{r()}} (0,2,4) \rightarrow \cdots (5,7,4) 
\rightarrow_{\textsc{r()}}(0,4,3)\rightarrow (5,4,3) 
\rightarrow_{\textsc{r()}}(0,1,2)\rightarrow \cdots (5,6,2) 
\rightarrow_{\textsc{r()}}(0,3,1)\rightarrow (5,3,1) 
\rightarrow_{\textsc{r()}}(0,0,0)$.
Similarly, the sequence of counter values $(r_1,r_2)$ at each reduce step
is $(2,0),(3,-5),(1,0),(2,-5),(0,0)$.

Now we prove by induction on the number of steps that 
\begin{equation} \label{r_j}
r_j = \frac{c_1 a_{j,1} + \cdots + c_n a_{j,n}}{s_{l}}.
\end{equation}
Trivially the property holds at the beginning as all counters $c_i$ are set to $0$.

If the $k$-th step is \textsc{increase($i$)}, this new value will be:
\[ r_j + a_{j,i} = r_j + \frac{s_l}{s_l} a_{j,i} = \frac{c_1 a_{j,1} + \cdots + (c_i + s_l) a_{j,i} + \cdots + c_n a_{j,n}}{s_{l}} =  r'_j.
\]
If the $k$-th step is \textsc{reduce()}, then:
\[ r'_j = \frac{(c_1-s_1) a_{j,1} + \cdots + (c_n-s_n) a_{j,n}}{s_l} = r_j - \frac{s_1 a_{j,1} + \cdots +s_n a_{j,n}}{s_{l}} = r_j
\]
(note that \textsc{reduce()} steps do not affect the counters $r_j$).

This proves expression (\ref{r_j}). Furthermore, 
since $\frac{c_i}{s_l} < 2$, we have:
\[ |r_j| = |\frac{c_1 a_{j,1}}{s_l} + \cdots + \frac{c_n a_{j,n}}{s_l}| < 2 \cdot n \cdot max_i|a_{j,i}|
\]
which gives an upper bound on the absolute value of the 
counters $r_1, \dots, r_m$. 

We define now the partition $\{V^k\}_{k \in [1,t]}$ (where $t$ is
the number of \textsc{reduce} steps) by taking as 
$V^1$ a set of vertices containing exactly 
one vertex labelled by each $i \in [1,n]$ 
and as $V^k$ (for $k>1$) a set of vertices containing exactly 
one vertex labelled by $i$
for
each \textsc{increase($i$)} operation done
between the $k$-th and $(k+1)$-th \textsc{reduce} step. 

Now, it remains to define the partition of edges $\{E^k\}_{k \in [1,t]}$.
First we define for each vertex $v$ labelled by $i$ of the set 
$V^1\cup\dots\cup V^k$ ($k \geq 1$) 
and each constraint $j$ the number of \emph{open} edges.
Let $\mathit{open}_{j,i}(v) = |a_{j,i}| - |\{(v,v')\in E_j\cap (E^1 \cup \dots \cup E^k)\}|$.
Then, $\mathit{open}_{j,i}(V^1\cup\dots\cup V^k) = \sum_{v \in (V^1\cup\dots\cup V^k)\cap V_i}
\mathit{open}_{j,i}(v)$.
We will show that the number of open edges $\mathit{open}_{j,i}(V^1\cup\dots\cup V^k)$
can be bounded by $|a_{j,i}|$ for each $k$.
That means that each subgraph
$(V^1\cup\dots\cup V^k,E^1\cup\dots\cup E^k)$ contains at most one vertex
labelled by $i$ not completely matched. That in turn means that
$B_p^{k}$ never contains more than $2$ vertices
labelled by $i$, since $B_p^{k}$ is composed of all vertices of
$V^1\cup\dots\cup V^k$ not completely matched as well as all vertices
of $V^k$ (which contains at most one vertex for each variable).

We first define, from the sequence of values $c_j^1,\ldots,c_j^t$ of 
$c_j$ after each reduce step
for each variable $i$, a sequence $c^1_{j,i},\ldots,c^t_{j,i}$ of integers.
These integers will indicate the number of open edges for each type of vertex
after each reduce step (the number is positive or negative depending
on the sign of $a_{j,i}$).
Let $r_j^1,\ldots,r_j^t$ be the sequence of values of the counter $r_j$ after
reduce steps.
We define the sets $\mathit{pos}_j = \{i\;|\;a_{j,i}\geq 0\}$ 
and $\mathit{neg}_j=\{i\;|\;a_{j,i}<0\}$.
Then, for each $k \in [1,t]$ we define for each value $r_j^k$ its positive part $r^k_{j,\mathit{pos}} = 
(\sum_{p \in \mathit{pos}_j} (c_p^k \cdot a_{j,p}))/s_{l}$ and its negative part
$r^k_{j,\mathit{neg}} = 
(\sum_{p \in \mathit{neg}_j} (c_p^k \cdot a_{j,p}))/{s_{l}}$ such that $r_j^k = r^k_{j,\mathit{pos}} + r^k_{j,\mathit{neg}}$.
In the example we have the following successive values for the $(r^k_{1,\mathit{pos}},r^k_{2,\mathit{pos}})$~: $(2,\frac{4}{5}),(3,\frac{3}{5}),(1,\frac{2}{5}),(2,\frac{1}{5}),
(0,0)$ and
for $(r^k_{1,\mathit{neg}},r^k_{2,\mathit{neg}})$~: 
$(0,-\frac{4}{5}),(0,-\frac{8}{5}),(0,-\frac{2}{5}),(0,-\frac{6}{5}),(0,0)$.
Now, it is easy to see that 
we can choose $c^k_{j,i} \in \{\lfloor \frac{a_{j,i}c_i^k}{s_l} \rfloor,
\lceil  \frac{a_{j,i}c_i^k}{s_l}\rceil\}$ such that
(a) $\sum_{p \in \mathit{pos}_j} c^k_{j,p} =  \lceil r^k_{j,\mathit{pos}} \rceil$,
(b) $\sum_{p \in \mathit{neg}_j} c^k_{j,p} =  \lfloor r^k_{j,\mathit{neg}} \rfloor$
and (c) $|c_{j,i}^k| \geq |c_{j,i}^{k+1}|$, if there was no \textsc{increase($i$)}
operation between the $k$-th and the $(k+1)$-th \textsc{reduce()}.
(a) and (b) guarantee $c^k_{j,1}+\cdots+c^k_{j,n}=r^k_j$. 
Furthermore, we have $|c^k_{j,i}| \leq |a_{j,i}|$, as $0 \leq c_{j}^k \leq s_l$.
In the example we choose as successive values for $(c^k_{1,1},c^k_{1,2},c^k_{1,3})$~:
$(0,2,0),$ $(0,3,0),$ $(0,1,0),$ $(0,2,0),$ $(0,0,0)$ and we choose
as successive values for $(c^k_{2,1},c^k_{2,2},c^k_{2,3})$~: 
$(0,-1,1),$ $(0,-2,1),$ $(0,-1,1),$ $(0,-2,1),$ $(0,0,0)$.

Now, we can show that we can choose
$\{E^k\}_{k \in [1,t]}$ such that $\mathit{open}_{j,i}(V^1\cup\dots\cup V^k)=|c^k_{j,i}|$.
Furthermore, since $|c^k_{j,i}| \leq |a_{j,i}|$
we can always make sure that there is at most one
not fully matched vertex for each variable $i$ in $V^1\cup\dots\cup V^k$.
To show that inductively 
let us consider the situation just before the $k$-th \textsc{reduce()}
step. $V_k$ contains vertices corresponding to variables $i$
with an \textsc{increase($i$)} operation after the $(k-1)$-th \textsc{reduce()}
(for $k=1$, $V_k$ contains a vertex for each variable $i$).
The number of open edges (before adding $E^k$) of variable $i$ which we call
$d_{j,i}^{k-1}$ is given by
$d^{k-1}_{j,i} = c^{k-1}_{j,i}+a_{j,i}$ (or just $a_{j,i}$ for $k=1$) for
the vertices labelled by variable $i$ for which an \textsc{increase($i$)} operation
has been performed after the $(k-1)$-th \textsc{reduce()};
and the number of open edges is $d^{k-1}_{j,i}=c^{k-1}_{j,i}$ for the other variables $i$.
We know that 
$\sum_{p \in \mathit{pos}_j} d^{k-1}_{j,p} + \sum_{p \in \mathit{neg}_j} d^{k-1}_{j,p}$ 
is equal to 
$\sum_{p \in \mathit{pos}_j} c^{k}_{j,p} + \sum_{p \in \mathit{neg}_j} c^{k}_{j,p}$
because of (a) and (b). That means that before and after a \textsc{reduce()}
the difference between ``positive'' and ``negative'' open edges is the same.
Furthermore $\sum_{p \in \mathit{pos}_j} d^{k-1}_{j,p} \geq \sum_{p \in \mathit{pos}_j} c^{k}_{j,p}$
and  $\sum_{p \in \mathit{neg}_j} d^{k-1}_{j,p} \leq \sum_{p \in \mathit{neg}_j} c^{k}_{j,p}$
and due to (c), $|c^{k}_{j,i}|$ decreases w.r.t. $|c^{k-1}_{j,i}|$ for 
not increased variables. Therefore, 
$E^k$ can be defined such that the number of open ``positive'' edges
and open ``negative'' edges decreases simultaneously to get
to $c^k_{j,i}$ from $d^{k-1}_{j,i}$.
This concludes the proof for $\vec{b} = \vec{0}$.

If $\vec{b} \not= \vec{0}$, we just consider having an additional variable
labelled by $0$ with coefficients $a_{j,0} = -b_j$ (for $1 \leq j \leq m$).
The vertex labelled by $0$ can be put into all $V^k$.
The edges involving $0$ are computed like the other edges.
\end{proof}

From Lemma \ref{lem:pw} and Proposition \ref{lemmaIII} we get the following
theorem.

\begin{theorem}\label{thm:pw}
An ILP instance $A\vec{x}=\vec{b}$ with $n$ variables is 
feasible if and only if 
there exists a graph $G \in {\mathcal G}[A\vec{x}=\vec{b}]$ 
of path-width bounded by $2n$.
\end{theorem}

From that we obtain the following corollary.

\begin{corollary}\label{cor:pw}
An ILP instance $A\vec{x}=\vec{b}$ with $n$ variables is
feasible if and only if the first order formula $\Phi[A\vec{x}=\vec{b}]$ is
satisfiable 
on the class of graphs with path-width $2n$.
\end{corollary}

\section{Automata construction for ILP}\label{automata}
In this section, we show a direct automata construction from an ILP instance $A\vec{x}=\vec{b}$ such that the Parikh image of the automaton coincides with the set of solutions of the ILP instance. We call such machines ILP automata.

We reuse the ideas in the proof of Lemma~\ref{lem:pw} from Section \ref{boundedpathwidth} in order to build an automaton whose states are tuples of integer numbers representing the possible values of the counters $r_1, \ldots, r_m$,  paired with a bit $B\in\{0,1\}$. 
%
%
We can think of each accepting run of the automaton as a way of discovering a solution $(x_1 = s_1,\ldots, x_n = s_n)$ for the ILP instance, starting with an initial assignment $(x_1 = 0,\ldots, x_n = 0)$ and continuing by increasing exactly one $x_i$ at each step. The run should also contain a step where the vector of coefficients $-\vec{b}$ is added to the current valuation of $r_1, \ldots, r_m$. The bit $B$ is used to ensure that $-\vec{b}$ is added exactly once.
The way we have enumerated graph vertices in order to obtain path decompositions of bounded width defines also the manner in which to pick an $x_i$ for the next increase such that the counters have bounded range (which implies that the state space is bounded). 
Formally, 

\begin{definition}[{\sc ILP Automata}]\label{ILPautomaton} Let $I \triangleq A\vec{x}=\vec{b}$ be an ILP instance over the variables $V=\{x_1,\ldots,x_n\}$ and $m$ constraints. The {\em ILP automaton} ${\cal A}_I$ associated to $I$ is the DFA $(\Sigma, Q, \delta, s_0, F)$ defined as follows. For a tuple of natural numbers $\vec{r}=(r_1, r_2,\ldots, r_m )$ we say that $\vec{r}$ is {\em bounded} iff $r_j\leq 2 \cdot n \cdot max_i|a_{j,i}|$ for every $j\in[1,m]$. Let $R_m$ be the set of all bounded $m$-tuples $\vec{r}$. Then,
\begin{itemize}
\item $\Sigma=V\cup\{b\}$ is the alphabet of ${\cal A}_I$;
\item $Q=(\{0,1\}\times R_m)$ is the set of states;
\item the transition map $\delta:Q\times \Sigma\mapsto 2^Q$ is defined as follows. Let $A_i$ be the $i$'th column of $A$. Then,
\[
  \delta((B,\vec{r}),x) = \left\{ 
  \begin{array}{l l}
    \{(B,\vec{r'})\mid \vec{r'}=\vec{r}+A_i\mbox{, $\vec{r'}$ is bounded}\} & \quad \text{if $x=x_i$ with $i\in[1,n]$};\\
    \{(1,\vec{r'})\mid \vec{r'}=\vec{r}-\vec{b}\mbox{, $\vec{r'}$ is bounded}\} & \quad \text{if $x=b$ and $B=0$};\\
    \emptyset & \quad \text{otherwise.}\\
  \end{array} \right.
\]

\item the initial state $s_0$ is the pair $(0,\vec{0}_m)$, where $\vec{0}_m$
is an $m$-tuple of $0$'s.

\item the set of final states $F$ is the singleton $\{(1,\vec{0}_m)\}$.\qed
\end{itemize}
\end{definition}

\medskip

Let $\Sigma=\{a_1,a_2,\ldots, a_t\}$ be an alphabet, and $\Sigma^*$ be the set of all words over $\Sigma$. The \emph{Parikh image} of a language $L\subseteq \Sigma^*$ is a mapping $\mathit{Parikh}:L \mapsto \mathbb{N}^t$ that associates to each word $w\in L$ the tuple of natural numbers $(p_1,p_2,\ldots,p_t)$,  where $p_i$ is the number of occurrences of the symbol $a_i$ in $w$, for every $i\in[1,t]$. 

\medskip

\begin{theorem} 
For any ILP instance $I \triangleq A\vec{x}=\vec{b}$, $Parikh(L({\cal A}_I)) = S_I$, where $L({\cal A}_I)$ is the language of ${\cal A}_I$ and $S_I\subseteq \mathbb{N}^n$ is the set of solutions of $I$.
\end{theorem}
\begin{proof}(Sketch) By the construction of ${\cal A}_I$, the Parikh image of any word accepted by the automaton is a solution of $I$. Now, given a solution $\vec{s}$ of $I$, take the sequence of steps {\sc increase}$(i)$ and {\sc reduce}$()$ used to define a path decomposition for the graph representation of $\vec{s}$ in Lemma~\ref{lem:pw}. The projection of this sequence on the steps {\sc increase}$(i)$ corresponds to an accepting run in the automaton ${\cal A}_I$ (each {\sc increase}$(i)$ corresponds to a transition over the symbol $x_i$).
\end{proof}

An interesting aspect of the automaton ${\cal A}_I$ is that it can be {\em implemented} as a compact Boolean program $P_I$ whose size is linear in the size of $I$, as opposed to the exponential size of ${\cal A}_I$.  $P_I$ has a (bounded) variable $r_i$ for each constraint, and a bit $B$ to keep track of whether $\vec{b}$ has already been used. These variables are all initialized to zero. $P_I$ iteratively guesses a symbol in $\Sigma$ and updates the variables according to the transition function of ${\cal A}_I$. Now a special control location is reachable if and only if ${\cal A}_I$ accepts a word (when all constraint counters are $0$ and $B$ is set to 1). The  intrinsic characteristic of $P_I$ is that checking the reachability of the special location gives an answer to the ILP problem, and further this can be done with any  
verification tools designed  for (Boolean) programs.


\section{Conclusion}\label{conclusions}
In this paper we have investigated whether the intuition of interpreting  ILP solutions with labelled graphs that are MSO definable and of bounded tree-width also applies to the ILP feasibility problem. We have given a positive answer to this question showing that ILP feasibility can indeed be reduced in polynomial time to the satisfiability problem of FO (rather than MSO) on the class of bounded path-width (as opposed to bounded tree-width) graphs which is again decidable by Seese's theorem \cite{Seese91}. What we have not explored yet is whether our approach could also entail the optimal complexity of the problem. Although the ILP feasibility problem is NP-complete, the Boolean programs derived from the automata construction of Section~\ref{automata} only lead to a PSPACE procedure. We believe it is interesting to shed some light in this regards. Furthermore, 
continuing the exploration in other directions by applying the approach of \cite{tree-width-popl} and the one we propose in this paper to other decision problems is also an interesting venue for future research. For example, for several other classes of automata their decision procedures for the emptiness problem is derived by checking that the Parikh image of the language accepted by them satisfies a set of linear constraints (see for example \cite{javierLICS12}). We believe that combining the behaviour graphs of these automata with the solution graphs we proposed for ILP could lead to further applications of the approach to broader classes of automata.

\bibliographystyle{eptcs}
\bibliography{ilp}

\begin{thebibliography}{10}
\providecommand{\bibitemdeclare}[2]{}
\providecommand{\surnamestart}{}
\providecommand{\surnameend}{}
\providecommand{\urlprefix}{Available at }
\providecommand{\url}[1]{\texttt{#1}}
\providecommand{\href}[2]{\texttt{#2}}
\providecommand{\urlalt}[2]{\href{#1}{#2}}
\providecommand{\doi}[1]{doi:\urlalt{http://dx.doi.org/#1}{#1}}
\providecommand{\bibinfo}[2]{#2}

\bibitemdeclare{article}{Alur}
\bibitem{Alur}
\bibinfo{author}{Rajeev \surnamestart Alur\surnameend} \&
  \bibinfo{author}{P.~\surnamestart Madhusudan\surnameend}
  (\bibinfo{year}{2009}): \emph{\bibinfo{title}{Adding nesting structure to
  words}}.
\newblock {\sl \bibinfo{journal}{J. ACM}}
  \bibinfo{volume}{56}(\bibinfo{number}{3}).
\newblock \urlprefix\url{http://doi.acm.org/10.1145/1516512.1516518}.

\bibitemdeclare{article}{faouziORDERED}
\bibitem{faouziORDERED}
\bibinfo{author}{Mohamed~Faouzi \surnamestart Atig\surnameend}
  (\bibinfo{year}{2012}): \emph{\bibinfo{title}{Model-Checking of Ordered
  Multi-Pushdown Automata}}.
\newblock {\sl \bibinfo{journal}{Logical Methods in Computer Science}}
  \bibinfo{volume}{8}(\bibinfo{number}{3}).
\newblock \urlprefix\url{http://dx.doi.org/10.2168/LMCS-8(3:20)2012}.

\bibitemdeclare{inproceedings}{ahmedLTL}
\bibitem{ahmedLTL}
\bibinfo{author}{Mohamed~Faouzi \surnamestart Atig\surnameend},
  \bibinfo{author}{Ahmed \surnamestart Bouajjani\surnameend},
  \bibinfo{author}{K.~Narayan \surnamestart Kumar\surnameend} \&
  \bibinfo{author}{Prakash \surnamestart Saivasan\surnameend}
  (\bibinfo{year}{2012}): \emph{\bibinfo{title}{Linear-Time Model-Checking for
  Multithreaded Programs under Scope-Bounding}}.
\newblock In \bibinfo{editor}{Supratik \surnamestart Chakraborty\surnameend} \&
  \bibinfo{editor}{Madhavan \surnamestart Mukund\surnameend}, editors: {\sl
  \bibinfo{booktitle}{ATVA}}, {\sl \bibinfo{series}{Lecture Notes in Computer
  Science}} \bibinfo{volume}{7561}, \bibinfo{publisher}{Springer}, pp.
  \bibinfo{pages}{152--166}.
\newblock \urlprefix\url{http://dx.doi.org/10.1007/978-3-642-33386-6_13}.

\bibitemdeclare{article}{Bodlaender}
\bibitem{Bodlaender}
\bibinfo{author}{Hans~L. \surnamestart Bodlaender\surnameend}
  (\bibinfo{year}{1993}): \emph{\bibinfo{title}{A Tourist Guide through
  Treewidth}}.
\newblock {\sl \bibinfo{journal}{Acta Cybern.}}
  \bibinfo{volume}{11}(\bibinfo{number}{1-2}), pp. \bibinfo{pages}{1--22}.
\newblock
  \urlprefix\url{http://www.inf.u-szeged.hu/kutatas/actacybernetica/vol11n12/b%
odlaen/bodlaen.xml}.

\bibitemdeclare{article}{Buchi60}
\bibitem{Buchi60}
\bibinfo{author}{J.~\surnamestart B{\"{u}}chi\surnameend}
  (\bibinfo{year}{1960}): \emph{\bibinfo{title}{Weak second-order Arithmetic
  and Finite Automata.}}
\newblock {\sl \bibinfo{journal}{Zeitschrift fur Mathematische Logik und
  Grundlagen der Mathematik}} \bibinfo{volume}{6}, pp. \bibinfo{pages}{66--92},
  \doi{10.1002/malq.19600060105}.

\bibitemdeclare{article}{courcelle}
\bibitem{courcelle}
\bibinfo{author}{Bruno \surnamestart Courcelle\surnameend}
  (\bibinfo{year}{1990}): \emph{\bibinfo{title}{The Monadic Second-Order Logic
  of Graphs. I. Recognizable Sets of Finite Graphs}}.
\newblock {\sl \bibinfo{journal}{Inf. Comput.}}
  \bibinfo{volume}{85}(\bibinfo{number}{1}), pp. \bibinfo{pages}{12--75}.
\newblock \urlprefix\url{http://dx.doi.org/10.1016/0890-5401(90)90043-H}.

\bibitemdeclare{inproceedings}{javierLICS12}
\bibitem{javierLICS12}
\bibinfo{author}{Javier \surnamestart Esparza\surnameend},
  \bibinfo{author}{Pierre \surnamestart Ganty\surnameend} \&
  \bibinfo{author}{Rupak \surnamestart Majumdar\surnameend}
  (\bibinfo{year}{2012}): \emph{\bibinfo{title}{A Perfect Model for Bounded
  Verification}}.
\newblock In: {\sl \bibinfo{booktitle}{LICS}}, \bibinfo{publisher}{IEEE}, pp.
  \bibinfo{pages}{285--294}.
\newblock \urlprefix\url{http://dx.doi.org/10.1109/LICS.2012.39}.

\bibitemdeclare{inproceedings}{Ganesh02}
\bibitem{Ganesh02}
\bibinfo{author}{Vijay \surnamestart Ganesh\surnameend},
  \bibinfo{author}{Sergey \surnamestart Berezin\surnameend} \&
  \bibinfo{author}{David~L. \surnamestart Dill\surnameend}
  (\bibinfo{year}{2002}): \emph{\bibinfo{title}{Deciding Presburger Arithmetic
  by Model Checking and Comparisons with Other Methods}}.
\newblock In \bibinfo{editor}{Mark \surnamestart Aagaard\surnameend} \&
  \bibinfo{editor}{John~W. \surnamestart O'Leary\surnameend}, editors: {\sl
  \bibinfo{booktitle}{FMCAD}}, {\sl \bibinfo{series}{Lecture Notes in Computer
  Science}} \bibinfo{volume}{2517}, \bibinfo{publisher}{Springer}, pp.
  \bibinfo{pages}{171--186}.
\newblock \urlprefix\url{http://dx.doi.org/10.1007/3-540-36126-X_11}.

\bibitemdeclare{techreport}{Gomory60}
\bibitem{Gomory60}
\bibinfo{author}{Ralph~E. \surnamestart Gomory\surnameend}
  (\bibinfo{year}{1960}): \emph{\bibinfo{title}{An algorithm for the mixed
  integer problem}}.
\newblock \bibinfo{type}{Technical Report}, \bibinfo{institution}{RAND
  Corporation}.

\bibitemdeclare{article}{Heussner}
\bibitem{Heussner}
\bibinfo{author}{Alexander \surnamestart Heu{\ss}ner\surnameend}
  (\bibinfo{year}{2012}): \emph{\bibinfo{title}{Model Checking Communicating
  Processes: Run Graphs, Graph Grammars, and MSO}}.
\newblock {\sl \bibinfo{journal}{ECEASST}} \bibinfo{volume}{47}.
\newblock
  \urlprefix\url{http://journal.ub.tu-berlin.de/eceasst/article/view/725}.

\bibitemdeclare{article}{muschollQUEUE}
\bibitem{muschollQUEUE}
\bibinfo{author}{Alexander \surnamestart Heu{\ss}ner\surnameend},
  \bibinfo{author}{J{\'e}r{\^o}me \surnamestart Leroux\surnameend},
  \bibinfo{author}{Anca \surnamestart Muscholl\surnameend} \&
  \bibinfo{author}{Gr{\'e}goire \surnamestart Sutre\surnameend}
  (\bibinfo{year}{2012}): \emph{\bibinfo{title}{Reachability Analysis of
  Communicating Pushdown Systems}}.
\newblock {\sl \bibinfo{journal}{Logical Methods in Computer Science}}
  \bibinfo{volume}{8}(\bibinfo{number}{3}).
\newblock \urlprefix\url{http://dx.doi.org/10.2168/LMCS-8(3:23)2012}.

\bibitemdeclare{inproceedings}{phase}
\bibitem{phase}
\bibinfo{author}{Salvatore \surnamestart {La Torre}\surnameend},
  \bibinfo{author}{P.~\surnamestart Madhusudan\surnameend} \&
  \bibinfo{author}{Gennaro \surnamestart Parlato\surnameend}
  (\bibinfo{year}{2007}): \emph{\bibinfo{title}{A Robust Class of
  Context-Sensitive Languages}}.
\newblock In: {\sl \bibinfo{booktitle}{LICS}}, \bibinfo{publisher}{IEEE
  Computer Society}, pp. \bibinfo{pages}{161--170}.
\newblock
  \urlprefix\url{http://doi.ieeecomputersociety.org/10.1109/LICS.2007.9}.

\bibitemdeclare{inproceedings}{gennaroTACAS08}
\bibitem{gennaroTACAS08}
\bibinfo{author}{Salvatore \surnamestart {La Torre}\surnameend},
  \bibinfo{author}{P.~\surnamestart Madhusudan\surnameend} \&
  \bibinfo{author}{Gennaro \surnamestart Parlato\surnameend}
  (\bibinfo{year}{2008}): \emph{\bibinfo{title}{Context-Bounded Analysis of
  Concurrent Queue Systems}}.
\newblock In \bibinfo{editor}{C.~R. \surnamestart Ramakrishnan\surnameend} \&
  \bibinfo{editor}{Jakob \surnamestart Rehof\surnameend}, editors: {\sl
  \bibinfo{booktitle}{TACAS}}, {\sl \bibinfo{series}{Lecture Notes in Computer
  Science}} \bibinfo{volume}{4963}, \bibinfo{publisher}{Springer}, pp.
  \bibinfo{pages}{299--314}.
\newblock \urlprefix\url{http://dx.doi.org/10.1007/978-3-540-78800-3_21}.

\bibitemdeclare{inproceedings}{salvatoreSCOPED}
\bibitem{salvatoreSCOPED}
\bibinfo{author}{Salvatore \surnamestart {La Torre}\surnameend} \&
  \bibinfo{author}{Margherita \surnamestart Napoli\surnameend}
  (\bibinfo{year}{2011}): \emph{\bibinfo{title}{Reachability of Multistack
  Pushdown Systems with Scope-Bounded Matching Relations}}.
\newblock In \bibinfo{editor}{Joost-Pieter \surnamestart Katoen\surnameend} \&
  \bibinfo{editor}{Barbara \surnamestart K{\"o}nig\surnameend}, editors: {\sl
  \bibinfo{booktitle}{CONCUR}}, {\sl \bibinfo{series}{Lecture Notes in Computer
  Science}} \bibinfo{volume}{6901}, \bibinfo{publisher}{Springer}, pp.
  \bibinfo{pages}{203--218}.
\newblock \urlprefix\url{http://dx.doi.org/10.1007/978-3-642-23217-6_14}.

\bibitemdeclare{inproceedings}{salvatoreLOGIC}
\bibitem{salvatoreLOGIC}
\bibinfo{author}{Salvatore \surnamestart {La Torre}\surnameend} \&
  \bibinfo{author}{Margherita \surnamestart Napoli\surnameend}
  (\bibinfo{year}{2012}): \emph{\bibinfo{title}{A Temporal Logic for
  Multi-threaded Programs}}.
\newblock In \bibinfo{editor}{Jos C.~M. \surnamestart Baeten\surnameend},
  \bibinfo{editor}{Thomas \surnamestart Ball\surnameend} \&
  \bibinfo{editor}{Frank~S. \surnamestart de~Boer\surnameend}, editors: {\sl
  \bibinfo{booktitle}{IFIP TCS}}, {\sl \bibinfo{series}{Lecture Notes in
  Computer Science}} \bibinfo{volume}{7604}, \bibinfo{publisher}{Springer}, pp.
  \bibinfo{pages}{225--239}.
\newblock \urlprefix\url{http://dx.doi.org/10.1007/978-3-642-33475-7_16}.

\bibitemdeclare{inproceedings}{scopedFSTTCS}
\bibitem{scopedFSTTCS}
\bibinfo{author}{Salvatore \surnamestart {La Torre}\surnameend} \&
  \bibinfo{author}{Gennaro \surnamestart Parlato\surnameend}
  (\bibinfo{year}{2012}): \emph{\bibinfo{title}{Scope-bounded Multistack
  Pushdown Systems: Fixed-Point, Sequentialization, and Tree-Width}}.
\newblock In \bibinfo{editor}{Deepak \surnamestart D'Souza\surnameend},
  \bibinfo{editor}{Telikepalli \surnamestart Kavitha\surnameend} \&
  \bibinfo{editor}{Jaikumar \surnamestart Radhakrishnan\surnameend}, editors:
  {\sl \bibinfo{booktitle}{FSTTCS}}, {\sl
  \bibinfo{series}{LIPIcs}}~\bibinfo{volume}{18}, \bibinfo{publisher}{Schloss
  Dagstuhl - Leibniz-Zentrum fuer Informatik}, pp. \bibinfo{pages}{173--184}.
\newblock \urlprefix\url{http://dx.doi.org/10.4230/LIPIcs.FSTTCS.2012.173}.

\bibitemdeclare{article}{Land60}
\bibitem{Land60}
\bibinfo{author}{A.~H. \surnamestart Land\surnameend} \& \bibinfo{author}{A.~G.
  \surnamestart Doig\surnameend} (\bibinfo{year}{1960}):
  \emph{\bibinfo{title}{An Automatic Method of Solving Discrete Programming
  Problems}}.
\newblock {\sl \bibinfo{journal}{Econometrica}}
  \bibinfo{volume}{28}(\bibinfo{number}{3}), pp. \bibinfo{pages}{497--520},
  \doi{10.2307/1910129}.

\bibitemdeclare{article}{LLL82}
\bibitem{LLL82}
\bibinfo{author}{A.~K. \surnamestart Lenstra\surnameend},
  \bibinfo{author}{H.~W.~Lenstra \surnamestart Jr.\surnameend} \&
  \bibinfo{author}{L.~\surnamestart Lovász\surnameend} (\bibinfo{year}{1982}):
  \emph{\bibinfo{title}{Factoring polynomials with rational coefficients}}.
\newblock {\sl \bibinfo{journal}{Mathematische Annalen}}
  \bibinfo{volume}{261}(\bibinfo{number}{4}), pp. \bibinfo{pages}{515--534},
  \doi{10.1007/BF01457454}.

\bibitemdeclare{inproceedings}{tree-width-popl}
\bibitem{tree-width-popl}
\bibinfo{author}{P.~\surnamestart Madhusudan\surnameend} \&
  \bibinfo{author}{Gennaro \surnamestart Parlato\surnameend}
  (\bibinfo{year}{2011}): \emph{\bibinfo{title}{The tree width of auxiliary
  storage}}.
\newblock In \bibinfo{editor}{Thomas \surnamestart Ball\surnameend} \&
  \bibinfo{editor}{Mooly \surnamestart Sagiv\surnameend}, editors: {\sl
  \bibinfo{booktitle}{POPL}}, \bibinfo{publisher}{ACM}, pp.
  \bibinfo{pages}{283--294}.
\newblock \urlprefix\url{http://doi.acm.org/10.1145/1926385.1926419}.

\bibitemdeclare{article}{Papadimitriou81}
\bibitem{Papadimitriou81}
\bibinfo{author}{Christos~H. \surnamestart Papadimitriou\surnameend}
  (\bibinfo{year}{1981}): \emph{\bibinfo{title}{On the complexity of integer
  programming}}.
\newblock {\sl \bibinfo{journal}{J. ACM}}
  \bibinfo{volume}{28}(\bibinfo{number}{4}), pp. \bibinfo{pages}{765--768}.
\newblock \urlprefix\url{http://doi.acm.org/10.1145/322276.322287}.

\bibitemdeclare{article}{parikh}
\bibitem{parikh}
\bibinfo{author}{Rohit \surnamestart Parikh\surnameend} (\bibinfo{year}{1966}):
  \emph{\bibinfo{title}{On Context-Free Languages}}.
\newblock {\sl \bibinfo{journal}{J. ACM}}
  \bibinfo{volume}{13}(\bibinfo{number}{4}), pp. \bibinfo{pages}{570--581}.
\newblock \urlprefix\url{http://doi.acm.org/10.1145/321356.321364}.

\bibitemdeclare{article}{Pugh92}
\bibitem{Pugh92}
\bibinfo{author}{William \surnamestart Pugh\surnameend} (\bibinfo{year}{1992}):
  \emph{\bibinfo{title}{A Practical Algorithm for Exact Array Dependence
  Analysis}}.
\newblock {\sl \bibinfo{journal}{Commun. ACM}}
  \bibinfo{volume}{35}(\bibinfo{number}{8}), pp. \bibinfo{pages}{102--114}.
\newblock \urlprefix\url{http://doi.acm.org/10.1145/135226.135233}.

\bibitemdeclare{inproceedings}{boundedCS}
\bibitem{boundedCS}
\bibinfo{author}{Shaz \surnamestart Qadeer\surnameend} \&
  \bibinfo{author}{Jakob \surnamestart Rehof\surnameend}
  (\bibinfo{year}{2005}): \emph{\bibinfo{title}{Context-Bounded Model Checking
  of Concurrent Software}}.
\newblock In \bibinfo{editor}{Nicolas \surnamestart Halbwachs\surnameend} \&
  \bibinfo{editor}{Lenore~D. \surnamestart Zuck\surnameend}, editors: {\sl
  \bibinfo{booktitle}{TACAS}}, {\sl \bibinfo{series}{Lecture Notes in Computer
  Science}} \bibinfo{volume}{3440}, \bibinfo{publisher}{Springer}, pp.
  \bibinfo{pages}{93--107}.
\newblock \urlprefix\url{http://dx.doi.org/10.1007/978-3-540-31980-1_7}.

\bibitemdeclare{article}{Seese91}
\bibitem{Seese91}
\bibinfo{author}{Detlef \surnamestart Seese\surnameend} (\bibinfo{year}{1991}):
  \emph{\bibinfo{title}{The Structure of Models of Decidable Monadic Theories
  of Graphs}}.
\newblock {\sl \bibinfo{journal}{Ann. Pure Appl. Logic}}
  \bibinfo{volume}{53}(\bibinfo{number}{2}), pp. \bibinfo{pages}{169--195}.
\newblock \urlprefix\url{http://dx.doi.org/10.1016/0168-0072(91)90054-P}.

\bibitemdeclare{inproceedings}{Wolper95}
\bibitem{Wolper95}
\bibinfo{author}{Pierre \surnamestart Wolper\surnameend} \&
  \bibinfo{author}{Bernard \surnamestart Boigelot\surnameend}
  (\bibinfo{year}{1995}): \emph{\bibinfo{title}{An Automata-Theoretic Approach
  to Presburger Arithmetic Constraints (Extended Abstract)}}.
\newblock In \bibinfo{editor}{Alan \surnamestart Mycroft\surnameend}, editor:
  {\sl \bibinfo{booktitle}{SAS}}, {\sl \bibinfo{series}{Lecture Notes in
  Computer Science}} \bibinfo{volume}{983}, \bibinfo{publisher}{Springer}, pp.
  \bibinfo{pages}{21--32}.
\newblock \urlprefix\url{http://dx.doi.org/10.1007/3-540-60360-3_30}.

\end{thebibliography}


\end{document}